\documentclass[reprint, superscriptaddress, twocolumn, nofootinbib, floatfix]{revtex4-1}
\usepackage[utf8]{inputenc}
\usepackage[lf,enc=t1]{berenis}
\usepackage[separate-uncertainty=true]{siunitx}
\usepackage{graphicx}
\usepackage{mathptmx}
\usepackage{amsmath}
\usepackage{amssymb}
\usepackage{etoolbox}
\usepackage{multirow}
\usepackage{dcolumn}
\usepackage{float}
\usepackage{tikz}
\usepackage{hyperref}
\DeclareMathOperator\erf{erf}

\definecolor{plot-red}{RGB}{255,0,0}
\definecolor{plot-blue}{RGB}{2,71,142}
\definecolor{plot-green}{RGB}{118,189,23}
\definecolor{plot-light-blue}{RGB}{0,176,240}
\makeatletter
\patchcmd{\@outputpage@head}{\@ifx{\LS@rot\@undefined}{}{\LS@rot}}{}{}{}
\makeatother

\begin{document}

\preprint{arxiv}

\title{Optimizing ToF-SIMS Depth Profiles of Semiconductor Heterostructures}

\author{Jan Tröger}
\affiliation{Institute of Materials Physics, University of Münster, 48149 Münster, Germany}
\affiliation{Tascon GmbH, 48149 Münster, Germany}
\email{jan.troeger@uni-muenster.de}
\author{Reinhard Kersting}
\affiliation{Tascon GmbH, 48149 Münster, Germany}
\author{Birgit Hagenhoff}
\affiliation{Tascon GmbH, 48149 Münster, Germany}
\author{Dominique Bougeard}
\affiliation{Institut für Experimentelle und Angewandte Physik, Universität Regensburg, 93040 Regensburg, Germany}
\author{Nikolay V. Abrosimov}
\affiliation{Leibniz-Institut für Kristallzüchtung (IKZ), 12489 Berlin, Germany}
\author{Jan Klos}
\affiliation{JARA-FIT Institute for Quantum Information, Forschungszentrum Jülich GmbH and RWTH Aachen University, 52056 Aachen, Germany}
\author{Lars R. Schreiber}
\affiliation{JARA-FIT Institute for Quantum Information, Forschungszentrum Jülich GmbH and RWTH Aachen University, 52056 Aachen, Germany}
\affiliation{ARQUE Systems GmbH, 52074 Aachen, Germany}
\author{Hartmut Bracht}
\affiliation{Institute of Materials Physics, University of Münster, 48149 Münster, Germany}

\begin{abstract}
The continuous technological development of electronic devices and the introduction of new materials leads to ever greater demands on the fabrication of semiconductor heterostructures and their characterization. This work focuses on optimizing Time-of-Flight Secondary Ion Mass Spectrometry (ToF-SIMS) depth profiles of semiconductor heterostructures aiming at a minimization of measurement-induced profile broadening. As model system, a state-of-the-art Molecular Beam Epitaxy (MBE) grown multilayer homostructure consisting of $^{\textit{nat}}$Si/$^{28}$Si bilayers with only \SI{2}{\nano\metre} in thickness is investigated while varying the most relevant sputter parameters. Atomic concentration-depth profiles are determined and an error function based description model is used to quantify layer thicknesses as well as profile broadening. The optimization process leads to an excellent resolution of the multilayer homostructure. The results of this optimization guide to a ToF-SIMS analysis of another MBE grown heterostructure consisting of a strained and highly purified $^{28}$Si layer sandwiched between two Si$_{0.7}$Ge$_{0.3}$ layers. The sandwiched $^{28}$Si layer represents a quantum well that has proven to be an excellent host for the implementation of electron-spin qubits.
\end{abstract}

\maketitle

\section{Introduction}
Semiconductor heterostructures, i.e. materials with semiconducting properties brought in direct contact with each other, are essential building blocks of modern electronic and optoelectronic devices. Combined into integrated circuits, they are the basis of almost all current information-processing devices, rendering applications such as solar cells, semiconductor lasers or light-emitting diodes possible \cite{Alferov2001, Kroemer2001, Kilby2001, Nakamura2015}. Furthermore, the realization of a quantum computer based on semiconductor heterostructures is also part of current research \cite{Xue2024, Burkard2023, Noiri2022, Philips2022, Xue2022, Huang2019, Hendrickx2021, Scappucci2021, Bluhm2019, Yoneda2018, Zwanenbourg2013, Wild2012}. In this advent of quantum technology applications and constantly progressing miniaturization, the demands on the precision of heterostructures are increasing \cite{Jmerik2022, Losert2023, Volmer2023, Klos2024}. The use of laser annealing, for example, is being discussed in order to avoid heating up the active region of the spin qubit devices during fabrication \cite{Neul2024} which leads to diffusional broadening of interfaces.

The technology development as well as fabrication control require proper analytical support. In particular, the determination of the material composition, the layer thickness and the intermixing between the individual layers are of prime importance, since they have a decisive effect on the performance \cite{Klos2024} and the service life of the devices. Several analytical methods are available for this task. A common technique is Scanning Transmission Electron Microscopy (STEM) imaging in combination with Energy Dispersive X-Ray Analysis (EDX) for the determination of the chemical composition. Based on the information carried by secondary electron contrasts, it offers high lateral resolution in the nanometer range and a sensitivity of about \SI{0.1}{\%} \cite{Servanton2009}. Another method that can be used for composition analysis, is Atom Probe Tomography (APT). Based on field ionization and evaporation of atoms from a needle-shaped specimen it offers an sub-\si{\nano\metre} resolution in three dimensions and a sensitivity of about \SI{0.1}{\%}\cite{Cerezo1986}. In the case sub-\si{\nano\metre} resolution in three dimensions is not required, a widely used technique for one-dimensional depth profiling is Secondary Ion Mass Spectrometry (SIMS). Primary ions are used for sputtering the sample which causes the emission of secondary ions from the sample. Using a Time-of-Flight mass spectrometer for mass analysis of the secondary ions (ToF-SIMS), depth resolutions of about \SI{1}{\nano\metre} and high sensitivities in the ppb-range can be achieved \cite{Moellers2019TOFSIMS5Pat}. Since no special preparation of the samples is necessary, SIMS is generally suitable for routine analysis of semiconductor heterostructures. The sputter process has already been widely studied \cite{Hofmann1998, Hofmann2000, Yamamura1982} and various approaches to minimize the sputter-related effects that limit the depth resolution by controlling individual measurement parameters have been presented \cite{Yamamura1988, Jahnel2003, Grehl2003, Liu2004}. However, the measurement conditions must be properly chosen for each material system to be studied in order to prevent the profile from broadening due to sputter-related effects such as atomic mixing and roughening of the sample surface. This work focuses on optimizing ToF-SIMS depth profiles of semiconductor heterostructures in favor of depth resolution by combining the most promising approaches. Using a test structure that consists of alternating natural Si and $^{28}$Si layers each \SI{2}{\nano\metre} in thickness, the influence of the main parameters is demonstrated and quantified. Then, a $^{28}$Si quantum well structure sandwiched between SiGe layers is analyzed by ToF-SIMS utilizing optimized instrumental settings that also consider measurement time and costs without significant loss of depth resolution. 

\section{Experimental} 
\begin{figure}
    \includegraphics{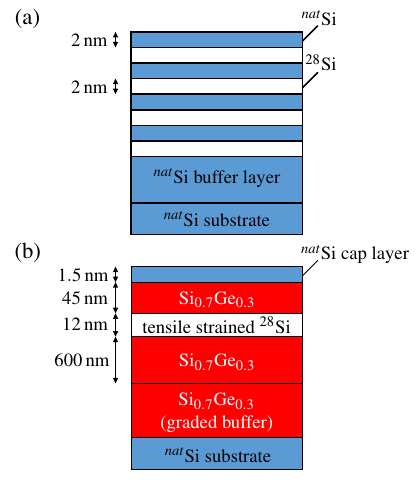}     
    \caption{Schematics of the analyzed samples. (a) sample \#1: a homostructure consisting of four $^{nat}$Si/$^{28}$Si bilayers was deposited onto a $^{nat}$Si buffer layer on top of a $^{nat}$Si substrate wafer by means of Molecular Beam Epitaxy (MBE). (b) sample \#2: a Si$_{0.7}$Ge$_{0.3}$/$^{28}$Si/Si$_{0.7}$Ge$_{0.3}$ heterostructure, protected from atmospheric oxygen by a $^{nat}$Si cap layer was deposited onto a strain-relaxed graded buffer up to a composition of Si$_{0.7}$Ge$_{0.3}$ on top of a $^{\textit{nat}}$Si substrate wafer. The highly purified and tensile strained $^{28}$Si layer represents an electron-spin quantum well.\cite{Struck2020}}
    \label{Fig:Samples_R2182_R2159}
\end{figure}

ToF-SIMS depth profiles were recorded on two different samples grown by means of Molecular Beam Epitaxy (MBE). Sample \#1 schematically shown in Fig. \ref{Fig:Samples_R2182_R2159} (a) was used as a test structure to study the influence of various instrumental parameters on profile broadening. It consists of four $^{\textit{nat}}$Si/$^{28}$Si bilayers that were deposited onto a $^{\textit{nat}}$Si buffer on top of a $^{\textit{nat}}$Si (001) substrate wafer. Every single layer of the test structure is nominally $\SI{2}{\nano\metre}$ thick. The sample is well suited for several reasons: First, silicon has been extensively studied and is the most common material for semiconductor heterostructures. On the other hand, by controlling the isotopic composition, a well-defined concentration profile of only a single chemical component is obtained. The latter is particularly advantageous since the same chemical properties of the sample components lead to mainly the same sputtering properties and matrix effects are therefore negligible.

Sample \#2 schematically shown in Fig. \ref{Fig:Samples_R2182_R2159} (b) represents a semiconductor heterostructure that consists of a highly purified and tensile strained $^{28}$Si layer sandwiched between two Si$_{0.7}$Ge$_{0.3}$ layers. The upper Si$_{0.7}$Ge$_{0.3}$ layer is protected from oxidation in air by a $^{\textit{nat}}$Si cap layer. The layer stack was deposited onto a strain-relaxed graded buffer up to a composition of Si$_{0.7}$Ge$_{0.3}$ on top of a $^{\textit{nat}}$Si substrate wafer \cite{Struck2020, Klos2024}. The nominal thicknesses are \SI{1.5}{\nano\metre} for the cap layer, \SI{45}{\nano\metre} for the upper Si$_{0.7}$Ge$_{0.3}$ layer, \SI{12}{\nano\metre} for the $^{28}$Si layer and \SI{600}{\nano\metre} for the lower Si$_{0.7}$Ge$_{0.3}$ layer of constant composition. The growth temperature was \SI{450}{\celsius}. In this configuration, the $^{28}$Si layer represents a quantum well, that is an excellent host for the implementation of electron-spin qubits \cite{Struck2024, Xue2022, Struck2020, Yoneda2018, Wild2012}.

On both samples, concentration depth profiles of the host elements were recorded using an IONTOF TOF.SIMS$^{5}$ instrument in dual-beam mode \cite{Iltgen1997, Grehl2003}. For sputtering, a low energy O$_2^{+}$ ion beam provided by a Dual Source Column (DSC) was used. A small area at the center of the sputter crater was analyzed using a dedicated high energy Bi$_x^{+}$ ion beam, provided by a Liquid Metal Ion Gun (LMIG). This approach offers the selection of sputtering and analysis conditions independent from each other.

In a variation study, the instrumental setting was optimized for minimal profile broadening when analyzing structures with only a few monolayers in thickness. Using sample \#1, several depth profiles were recorded while varying sputter and analysis parameters. All measurements carried out as part of the variation study are summarized in Tab. \ref{Tab:R2182_Measurements} together with the respective conditions. 
\begin{table*}
    \centering
    \caption{Overview of all SIMS analysis conditions applied for Si depth profiling of sample \#1. $X_{j}$: ion species, $E_{j}$: beam energy, $I_{j}$: beam current, $A_{j}$: rastered area, $p$: partial pressure of the flood gas, wherein $j=sp$ represents the sputter beam and $j=an$ the analysis beam. A $\uparrow$ symbol indicates an increase and a $\downarrow$ symbol indicates a decrease of the respective parameter.}
    \begin{ruledtabular}
    \begin{tabular}{lccdcccdcr} 
    \multicolumn{1}{l}{Variation}	& \multicolumn{1}{l}{$X_{sp}$}  	& \multicolumn{1}{l}{$E_{sp}$ ($\si{\electronvolt}$)}	& \multicolumn{1}{l}{$I_{sp}$ ($\si{\nano\ampere}$)} 	& \multicolumn{1}{l}{$A_{sp}$ ($\si{\micro\metre\squared}$)}	& \multicolumn{1}{l}{$X_{an}$}	& \multicolumn{1}{l}{$E_{an}$ ($\si{\kilo\electronvolt}$)}	& \multicolumn{1}{l}{$I_{an}$ ($\si{\pico\ampere}$)}	& \multicolumn{1}{l}{$A_{an}$ ($\si{\micro\metre\squared}$)}	& \multicolumn{1}{c}{$p$ ($\si{\milli\bar}$)} \\
    \hline
    Reference                   & O$_2^{+}$  & 500  & 81.0  & $300\times300$  & Bi$_1^{+}$	& 15 &  1.5 & $100\times100$  & \num{2E-6}  \\
    $p \uparrow$			    & O$_2^{+}$  & 500  & 81.0  & $300\times300$  & Bi$_1^{+}$ 	& 15 &  1.5 & $100\times100$  & \num{4E-6}  \\
    $p \downarrow$              & O$_2^{+}$  & 500  & 81.0  & $300\times300$  & Bi$_1^{+}$ 	& 15 &  1.5 & $100\times100$  & \num{5E-7}  \\
    $I_{an} \uparrow$           & O$_2^{+}$  & 500  & 81.0  & $300\times300$  & Bi$_1^{+}$	& 15 & 12.8 & $100\times100$  & \num{2E-6}  \\
    $I_{an} \downarrow$         & O$_2^{+}$  & 500  & 81.0  & $300\times300$  & Bi$_1^{+}$ 	& 15 &  0.2 & $100\times100$  & \num{2E-6}  \\
    $X_{an}$                    & O$_2^{+}$  & 500  & 81.0  & $300\times300$  & Bi$_3^{+}$ 	& 15 &  0.9 & $100\times100$  & \num{2E-6}  \\
    $E_{sp} \downarrow$		    & O$_2^{+}$  & 250  & 20.0  & $300\times300$  & Bi$_1^{+}$ 	& 15 &  1.5 & $100\times100$  & \num{2E-6}  \\
    $E_{sp}^{\ast} \downarrow$  & O$_2^{+}$  & 250  & 16.2  & $300\times300$  & Bi$_1^{+}$ 	& 13 &  1.5 &  $75\times75$   & \num{2E-6}  \\
    \end{tabular}
    \end{ruledtabular}
    \label{Tab:R2182_Measurements}
\end{table*}
For reference purposes, an instrumental setting used for routine analysis of semiconductor heterostructures with a few \si{\nano\metre} in thickness was applied first. Sputtering was performed using a O$_2^{+}$ ion beam with \SI{500}{\electronvolt} kinetic energy and a current of \SI{81}{\nano\ampere}, rastered over an area of $\SI{300}{\micro\metre} \times \SI{300}{\micro\metre}$. A Bi$_1^{+}$ ion beam with \SI{15}{\kilo\electronvolt} kinetic energy and a current of \SI{1.5}{\pico\ampere}, rastered over an area of $\SI{100}{\micro\metre} \times \SI{100}{\micro\metre}$ in the center of the sputter crater for analysis. The sample environment was flooded with oxygen gas in a controlled manner to ensure a completely oxidized sample surface throughout the measurement. This primarily avoids sputter-induced roughening that can occur due to different sputter rates of Si and its oxide and shortens the time to reach sputtering equilibrium. In addition, the presence of oxygen increases the ionization rate of cations, resulting in higher ion yields and therefore higher sensitivities. As reference, the partial pressure of oxygen was set to \SI{2E-06}{\milli\bar}. Secondary ions of positive polarity were analyzed. This setting represents a compromise between mass resolution, lateral resolution, depth resolution and measurement time. In a series of measurements the instrumental parameters were optimized in favour of depth resolution. First, the influence of the flood gas pressure, the primary ion current of the analysis beam and the primary ion species used for analysis were investigated. Finally, the sputtering conditions were optimized, with the energy of the sputter beam being the essential parameter. Changing one parameter was followed by the acquisition of a new depth profile, respectively, before the parameter was reset and another parameter was addressed. The sputter time was converted into a depth scale by measuring the depth of the sputter crater with an uncertainty of about \SI{5}{\%} using an IONTOF Atomic Force Microscope (AFM) based mechanical profilometer, which offers a wide scan range. The intensities were converted into a concentration scale referring to the isotopic distribution in $^{\textit{nat}}$Si \cite{Lide2009}. 

Depth profiling of the semiconductor heterostructure (sample \#2) was performed using a O$_2^{+}$ ion beam with \SI{500}{\electronvolt} kinetic energy and a maximum current of \SI{40}{\nano\ampere} for sputtering, rastered over an area of $\SI{300}{\micro\metre} \times \SI{300}{\micro\metre}$. A Bi$_1^{+}$ ion beam with \SI{15}{\kilo\electronvolt} kinetic energy and a current of \SI{1.5}{\pico\ampere} was rastered over an area of $\SI{75}{\micro\metre} \times \SI{75}{\micro\metre}$ in the center of the sputter crater for analysis. Oxygen gas flooding was used, with the partial pressure being \SI{2E-06}{\milli\bar}. Secondary ions of positive polarity were analyzed. The sputter time was converted into a depth scale by measuring the depth of the sputter crater with an uncertainty of about \SI{10}{\%} using an BRUKER DektakXT mechanical profilometer. The intensities were converted into a concentration scale considering the isotope distribution in the Si$_{0.7}$Ge$_{0.3}$ layers, which was determined based on the isotope distributions \cite{Lide2009} and the atomic densities \cite{Lide2009} of $^{\textit{nat}}$Si and $^{\textit{nat}}$Ge.

The broadening and the layer thickness were determined by fitting a model function that has been widely used for the analysis of self-atom profiles and its broadening in semiconductor heterostructures \cite{Paquelet2022, Dyck2017, Suedkamp2016, Kube2013, Kube2010}. Every interface was assumed to be described by an error function. Following this approach, the concentration profiles $C^{\ast}$ ($^{28}$Si, $^{29}$Si, $^{30}$Si) of sample \#1 are described by:
\begin{eqnarray}
	C^{\ast} && = \frac{C_1+C_3}{2} \nonumber\\
             && + \frac{C_1-C_2}{2} \sum_{i=1}^{7} (-1)^{i} \erf{\left( \frac{z-z_i}{r_i}\right)} \nonumber\\
             && + \frac{C_3-C_2}{2} \erf{\left(\frac{z-z_8}{r_8}\right)} \,,
\label{Eqn:R2182_Fit-Fkt}
\end{eqnarray}
where $C_1$ is the concentration of a Si isotope in the $^{nat}$Si layers, $C_2$ the concentration in the isotopically purified $^{28}$Si layers, $C_3$ the concentration in the $^{nat}$Si buffer layer, $r_i$ describes the broadening of interface $i$, counted starting at the sample surface, and $z_i$ with $z_{i+j}=z_i+z_{i+1}+...+z_{i+j-1}$ stands for the layer thickness. Fitting was done after the least square method, using $r_i$, $z_i$ and $C_2$ as free parameters. The concentration profiles $C^{\ast}$ ($^{28}$Si, $^{29}$Si, $^{30}$Si, $^{70}$Ge, $^{72}$Ge, $^{73}$Ge, $^{74}$Ge, $^{76}$Ge) of sample \#2 are described by:
\begin{eqnarray}
	C^{\ast} && = \frac{C_1+C_3}{2} \nonumber\\
             && - \frac{C_1-C_2}{2} \erf{\left( \frac{z-z_1}{r_1}\right)} \nonumber\\
             && + \frac{C_3-C_2}{2} \erf{\left(\frac{z-z_2}{r_2}\right)} \,,
    \label{Eqn:R2159_Fit-Fkt}
\end{eqnarray}
where $C_1$ is the concentration of a Si or Ge isotope in the upper Si$_{0.7}$Ge$_{0.3}$ layer, $C_2$ the concentration in the isotopically purified $^{28}$Si quantum well, $C_3$ the concentration in the lower Si$_{0.7}$Ge$_{0.3}$ layer, $r_i$ describes the broadening of interface $i$, counted starting at the sample surface, and $z_i$ with $z_{i+j}=z_i+z_{i+1}+...+z_{i+j-1}$ stands for the layer thickness, respectively. Fitting was done after the least square method, using $r_i$, $z_i$ and $C_2$ as free parameters.

A Park Systems XE-100 Atomic Force Microscope (AFM) was used for topography analysis on both samples (\#1 and \#2). Using a NSC15 tip an area of $\SI{8}{\micro\metre} \times \SI{8}{\micro\metre}$ was scanned in non-contact mode. AFM analysis was performed on the sample surface and in the crater resulting from the ToF-SIMS measurements.

\section{Results and Discussion}
\subsection{Depth profiling of a test homostructure - A parameter variation study}  
\begin{figure}
    \includegraphics{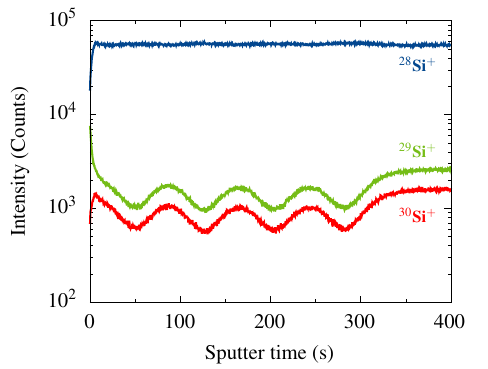} 
    \caption{SIMS depth profiles of $^{28}$Si, $^{29}$Si and $^{30}$Si in the four-bilayer $^{\text{nat}}$Si/$^{28}$Si test structure. Measurement conditions: O$_2^{+}$, \SI{500}{\electronvolt}, \SI{81}{\nano\ampere}, $\SI{300}{\micro\metre} \times \SI{300}{\micro\metre}$ sputtering; Bi$_1^{+}$, \SI{15}{\kilo\electronvolt}, \SI{1.5}{\pico\ampere}, $\SI{100}{\micro\metre} \times \SI{100}{\micro\metre}$ analysis; O$_2$, \SI{2E-6}{\milli\bar} flooding.}
    \label{Fig:R2182_RoutineConditions}
\end{figure}

\begin{figure}
    \includegraphics{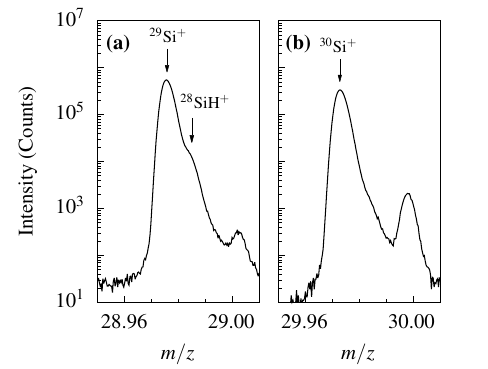} 
    \caption{Sections of the mass spectrum of the test structure. (a): on the right side of the $^{29}$Si peak an interference with $^{28}$SiH is observed. (b): for comparison the right side of the $^{30}$Si peak appears much steeper. A possible interference of $^{30}$Si and $^{29}$SiH can be neglected.}
    \label{Fig:R2182_MassSpectrum}
\end{figure}

The results of the ToF-SIMS reference measurement using the setting for routine analysis are shown in Fig. \ref{Fig:R2182_RoutineConditions}. 
Profiles of all three stable Si isotopes are shown. The measured $^{29}$Si and $^{30}$Si intensities show a clearly visible modulation. A similar modulation is observed for the measured $^{28}$Si intensity, but with smaller differences between local minima and maxima of the measured intensity compared to the $^{29}$Si and $^{30}$Si signals that are not visible on the logarithmic scale in Fig. \ref{Fig:R2182_RoutineConditions}. The reason are smaller differences in the $^{28}$Si concentration between the $^{\text{nat}}$Si layers and the purified $^{28}$Si layers in comparison to the $^{29}$Si concentration and the $^{30}$Si concentration resulting directly from the isotopic distribution of $^{\text{nat}}$Si. The onset of the $^{29}$Si signal suggests a high $^{29}$Si concentration within the first few \si{\nano\metre}. This is due to an interference of the $^{29}$Si signal with a $^{28}$SiH signal. In the mass spectrum, shown in Fig. \ref{Fig:R2182_MassSpectrum}, a $^{28}$SiH peak is observed in overlay with the $^{29}$Si peak (a), while the right flank of the $^{30}$Si peak (b) appears much steeper. Thus, a possible interference of $^{30}$Si and $^{29}$SiH can be neglected. Only the $^{30}$Si signal was hence considered for the parameter variation study.

\subsubsection{Influence of the flood gas pressure}
Using oxygen gas flooding with a partial pressure of \SI{2E-6}{\milli\bar}, the $^{30}$Si signal was detected with maximum intensity. When the partial pressure was increased to \SI{4E-6}{\milli\bar}, no significant influence on the measured profile was observed. In contrast, a reduction in the partial oxygen pressure lead to a clearly stronger profile broadening (not shown as a figure) followed by an increase in secondary ion intensity during the first few \si{\nano\metre}. This suggests an increasing concentration of $^{30}$Si within the upper most $^{\text{nat}}$Si layer. This is related to the reaction rate of the oxidation process. Below a partial pressure of \SI{2E-6}{\milli\bar}, the maximum reaction rate for the oxidation of the sample surface is not reached and implantation of oxygen into the sample surface during the sputter process leads to an increasing ion intensity, until the sputter equilibrium is reached, thereby suggesting an increased concentration of $^{30}$Si. A partial pressure of \SI{2E-6}{\milli\bar} that was already established for the reference measurement is considered to be sufficient to establish equilibrium conditions after short sputter times and accordingly was kept for the following measurements.

\subsubsection{Influence of the analysis beam current}
\begin{figure}
    \includegraphics{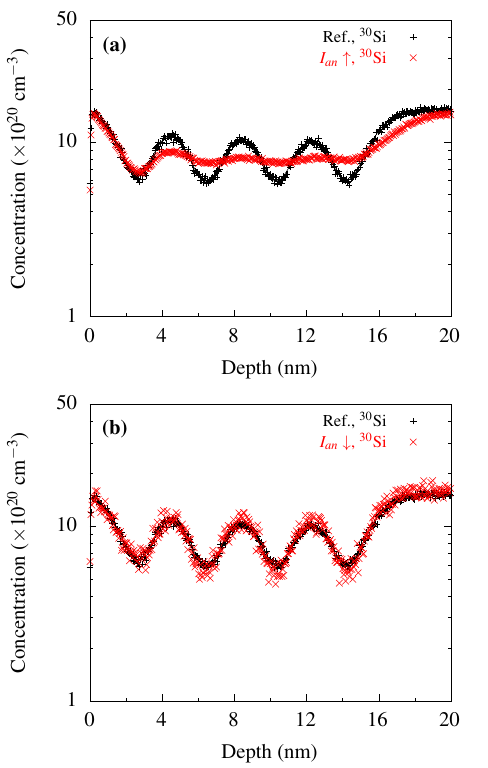} 
    \caption{SIMS concentration-depth profiles of $^{30}$Si in the four-bilayer $^{\text{nat}}$Si/$^{28}$Si test structure, recorded under different experimental conditions. Plus-signs mark data points recorded under reference conditions (O$_2^{+}$, \SI{500}{\electronvolt}, \SI{81}{\nano\ampere}, $\SI{300}{\micro\metre} \times \SI{300}{\micro\metre}$ sputtering; Bi$_1^{+}$, \SI{15}{\kilo\electronvolt}, \SI{1.5}{\pico\ampere}, $\SI{100}{\micro\metre} \times \SI{100}{\micro\metre}$ analysis; O$_2$, \SI{2E-6}{\milli\bar} flooding). Crosses mark data points recorded with the analysis beam current increased to \SI{12.8} {\pico\ampere} (a) and decreased to \SI{0.2}{\pico\ampere} (b) compared to the reference conditions, respectively. Every 3rd data point is shown for clarity. The solid lines represent the best fits based on Eq. \ref{Eqn:R2182_Fit-Fkt} taking sputter broadening effects into account.}
    \label{Fig:R2182_AnalysisCurrent}
\end{figure}

The ion current of the analysis beam has a noticeable influence on the profile: increasing the  ion current leads to higher intensities but lower dynamic range and visible profile broadening, which increases with depth as shown in Fig.\ref{Fig:R2182_AnalysisCurrent} (a). These effects are caused by the larger number of high-energy ion collisions, resulting in strong atomic mixing and damage of the samples structure that accumulates with depth. In contrast, decreasing the ion current extends the dynamic range and lowers the profile broadening, but also leads to stronger scattering of the data, as shown in Fig. \ref{Fig:R2182_AnalysisCurrent} (b). The accuracy of the data scales with $\sqrt{N}$ where $N$ describes the measured ion intensity (counts), i.e. higher intensities can be measured with greater accuracy than lower intensities. Reducing the primary ion current directly leads to lower secondary ion intensities, resulting in lower accuracy of the data. Therefore, decreasing the primary ion current does not improve the profile quality.

\subsubsection{Influence of the analysis beam species}
\begin{figure}
    \includegraphics{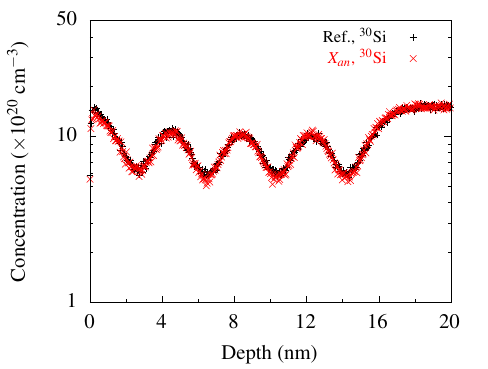} 
    \caption{SIMS concentration-depth profiles of $^{30}$Si in the four-bilayer $^{\text{nat}}$Si/$^{28}$Si test structure, recorded under different experimental conditions. Plus-signs mark data points recorded under reference conditions (O$_2^{+}$, \SI{500}{\electronvolt}, \SI{81}{\nano\ampere}, $\SI{300}{\micro\metre} \times \SI{300}{\micro\metre}$ sputtering; Bi$_1^{+}$, \SI{15}{\kilo\electronvolt}, \SI{1.5}{\pico\ampere}, $\SI{100}{\micro\metre} \times \SI{100}{\micro\metre}$ analysis; O$_2$, \SI{2E-6}{\milli\bar} flooding). Crosses mark data points recorded using a Bi$_3^{+}$ cluster ion beam with a current of \SI{0.9}{\pico\ampere} for analysis compared to the reference conditions. Every 3rd data point is shown for clarity. The solid lines represent the best fits based on Eq. \ref{Eqn:R2182_Fit-Fkt} taking sputter broadening effects into account.}
    \label{Fig:R2182_AnalysisSpecies}
\end{figure}

Theoretically, the use of Bi$_3^{+}$ instead of Bi$_1^{+}$ should be advantageous. Bi$_3^{+}$ is a cluster ion that dissociates when it hits the sample surface. Thus, only one-third of the total energy of the sputter beam is deposited by each of the three individual Bi atoms near the point of impact, resulting in a broader distribution of total energy. This should confine the bombardment-induced atomic mixing to a smaller volume. The extent of this confinement has been discussed in the SIMS community for many years \cite{Medvedeva2002}. Surprisingly, Fig. \ref{Fig:R2182_AnalysisSpecies} reveals no improvement of depth resolution by the use of Bi$_3^{+}$ cluster ions compared to the analysis under reference conditions.

\subsubsection{Influence of the sputter beam energy}
\begin{figure}
    \includegraphics{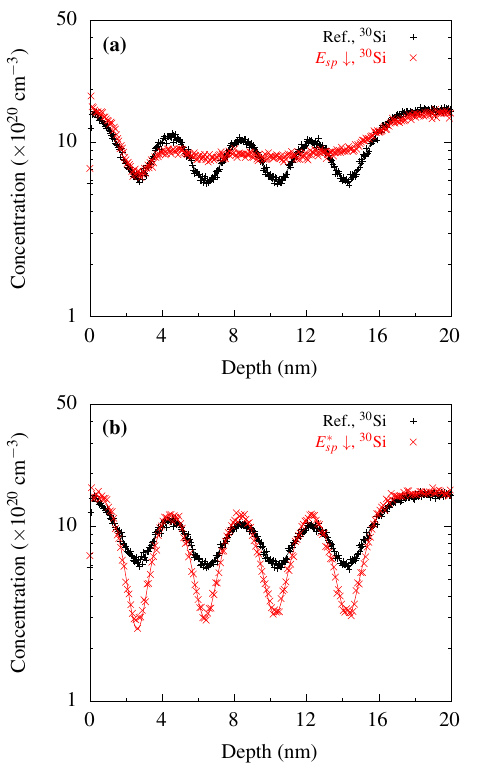} 
    \caption{SIMS concentration-depth profiles of $^{30}$Si in the four-bilayer $^{\text{nat}}$Si/$^{28}$Si test structure, recorded under different experimental conditions. Plus-signs mark data points recorded under reference conditions (O$_2^{+}$, \SI{500}{\electronvolt}, \SI{81}{\nano\ampere}, $\SI{300}{\micro\metre} \times \SI{300}{\micro\metre}$ sputtering; Bi$_1^{+}$, \SI{15}{\kilo\electronvolt}, \SI{1.5}{\pico\ampere}, $\SI{100}{\micro\metre} \times \SI{100}{\micro\metre}$ analysis; O$_2$, \SI{2E-6}{\milli\bar} flooding). Crosses mark data points recorded with a sputter beam energy decreased to \SI{250}{\electronvolt} reaching a maximum beam current of \SI{20}{\nano\ampere}. (a): the sputter rate ratio was not taken into account and calculated to be $47$. Every 3rd ($^{30}$Si, Ref.) and every 40th ($E_{sp} \downarrow$) data point is shown for clarity. (b): the sputter rate ratio was taken into account (Bi$_1^{+}$, \SI{13}{\kilo\electronvolt}, $\SI{75}{\micro\metre} \times \SI{75}{\micro\metre}$ analysis; sputter time between consecutive analysis cycles extended) and calculated to be $1233$. Every 3rd data point is shown for clarity. The solid lines represent best fits based on Eq. \ref{Eqn:R2182_Fit-Fkt} taking sputter broadening effects into account.}
    \label{Fig:R2182_SputterEnergy}
\end{figure}

It is well known that the energy of sputtering ions is crucial for depth resolution in sputter depth profiling and effects that lead to profile broadening generally scale with the sputtering energy. Therefore, reducing the sputtering energy is obvious while optimizing the analysis conditions for minimal profile broadening. In Fig. \ref{Fig:R2182_SputterEnergy} (a) a profile acquired reducing the sputtering energy from \SI{500}{\electronvolt} to \SI{250}{\electronvolt} is shown in comparison to the reference profile.
Instead of an improvement in profile quality, a lacking dynamic range and strong profile broadening that increases with depth is observed. The profile shows some similarities to the profile acquired after increasing the primary ion current (Fig. \ref{Fig:R2182_AnalysisCurrent} (a)) suggesting a similar reason: a reduction in sputtering energy primarily leads to significantly slower material removal. However, if all other parameters, in particular the number of analysis cycles between two sputtering cycles are kept constant, the high-energy analysis beam hits the sample much more often before the profile has been fully recorded. Note that the number of data points in the profile $E_{sp} \downarrow$ is significantly higher compared to the reference profile (Fig. \ref{Fig:R2182_SputterEnergy} (a)). The consequence of the higher dose of high-energy primary ions is structural damage of the sample. Therefore, the sputter rate ratio $R^{\ast}$ describing the amount of material removed by both, the sputter and analysis beam, has to be taken into account. It is defined by:
\begin{equation}
	R^{\ast} = \frac{R_{sp}}{R_{an}} \propto \frac{D_{sp}}{D_{an}} \cdot \frac{Y_{sp}}{Y_{an}} \cdot \frac{A_{an}}{A_{sp}} \, ,
	\label{Eq:SputterRateRatio}
\end{equation}
where $D_{x}$ is the ion dose, $Y_{x}$ the yield and $A_{x}$ the rastered area of the sputter beam ($x=sp$) and the analysis beam ($x=an$), respectively. According to experimental data \cite{Grehl2003PhD}, a limit value for significant loss of depth resolution is $R^{\ast} \approx 500$, while an improvement of depth resolution with increasing $R^{\ast}$ is observed up to $R^{\ast} \approx 1000$. Below the lower limit, the material removal is no longer dominated by the sputter beam and structural damage caused by the analysis beam can accumulate during the measurement. Ideally, the damage is removed within the next sputtering cycle. 

For the profile $E_{sp} \downarrow$ in Fig. \ref{Fig:R2182_SputterEnergy} (a) the sputter rate ratio $R^{\ast}$ was calculated to be $47$ which is well below the limit for significant loss of depth resolution. In order to successfully perform depth profiling with low energies, it must be ensured that the material removal of both ion beams is balanced in respect of $R^{\ast}$. Therefore, the sputtering time between successive analysis cycles was increased. In addition, the energy of the analysis beam was decreased to $\SI{13}{\kilo\electronvolt}$ and the analyzed area that also scales the sputter rate ratio ($A_{an}$ in Eq. \ref{Eq:SputterRateRatio}) was decreased to $\SI{75}{\micro\metre} \times \SI{75}{\micro\metre}$. This results in a sputter rate ratio of $1233$. In the profile, shown in Fig. \ref{Fig:R2182_SputterEnergy} (b), a larger dynamic range and significantly reduced instrumental broadening is observed.

For comparison, the sputter rate ratios were calculated for all recorded profiles and summarized in Tab \ref{Tab:R2182_SputterRateRatios_Broadening}.
\begin{table}
    \centering
    \caption{Sputter rate ratios $R^{\ast}$ and average changes in the interfacial broadening $\varnothing r_{i}$ compared to the reference profile. Positive values indicate an increase, while negative values indicate an improvement in interfacial broadening. $p$: partial pressure of the flood gas, $I_{j}$: beam current, $X_{j}$: ion species, $E_{j}$: beam energy, wherein $j=sp$ represents the sputter beam and $j=an$ the analysis beam. A $\uparrow$ symbol indicates an increase and a $\downarrow$ symbol indicates a decrease of the respective parameter. n.d.: value not determined.}
    \begin{ruledtabular}
    \begin{tabular}{llrrrr}
    \multicolumn{1}{l}{Variation} & \multicolumn{1}{l}{Fig.} & \multicolumn{1}{c}{$R^{\ast}$} & \multicolumn{1}{l}{$\varnothing r_{i}$} ($\si{\%}$) & \multicolumn{1}{l}{$\varnothing r_{i}$} ($\si{\%}$) & \multicolumn{1}{c}{$\varnothing r_{i}$ ($\si{\%}$)} \\
    &   &   & \multicolumn{1}{c}{$_{i=1,3,5,7}$}    & \multicolumn{1}{c}{$_{i=2,4,6,8}$}    & \multicolumn{1}{c}{$_{i=1,...,8}$}      \\ 
    \hline
    Reference                       & \ref{Fig:R2182_AnalysisCurrent} -- \ref{Fig:R2182_SputterEnergy}  &  $863$ & $0$    & $0$    & $0$    \\
    $p \downarrow$               & --                                 &  $863$ & $+33$  & $-9$   & $+12$  \\
    $p \uparrow$                 & --                                 &  $863$ & $-9$   & $+3$   & $-3$   \\
    $I_{an} \uparrow$            & \ref{Fig:R2182_AnalysisCurrent} (a) &  $101$ & n.d. & n.d. & n.d. \\
    $I_{an} \downarrow$          & \ref{Fig:R2182_AnalysisCurrent} (b) & $6471$ & $-6$   & $-9$   & $-8$   \\
    $X_{an}$                       & \ref{Fig:R2182_AnalysisSpecies}     &  $863$ & $-2$   & $-2$   & $-2$   \\
    $E_{sp} \downarrow$          & \ref{Fig:R2182_SputterEnergy} (a)   &   $47$ & n.d. & n.d. & n.d. \\
    $E_{sp}^{\ast} \downarrow$   & \ref{Fig:R2182_SputterEnergy} (b)   & $1233$ & $-28$  & $-26$  & $-27$  \\ 
    \end{tabular}
    \end{ruledtabular}
    \label{Tab:R2182_SputterRateRatios_Broadening}
\end{table}
In case of $I_{an} \uparrow$ and $E_{sp} \downarrow$ the limit of $R^{\ast} \approx 500$ was not reached. All further measurements revealed a sputter rate ratio $R^{\ast} > 500$. For the reference profile, as well as the profiles $p \downarrow$, $p \uparrow$, and $X_{an}$ the respective sputter rate ratios of $863$ are well above the lower limit of $R^{\ast} \approx 500$, but still below $R^{\ast} \approx 1000$ indicating room for optimization. The highest sputter rate ratio of $6471$ was determined for the profile $I_{an} \downarrow$, but the profile $E_{sp} ^{\ast} \downarrow$ that revealed a sputter rate ratio of $1233$ shows higher depth resolution (cf. Fig. \ref{Fig:R2182_AnalysisCurrent} (b) and Fig. \ref{Fig:R2182_SputterEnergy} (b)). This confirms that the sputtering rate ratio determines the depth resolution only up to $R^{\ast} \approx 1000$ and beyond the sputtering energy is dominant.

\subsubsection{Evaluation of the fit results}
\begin{figure}
    \includegraphics{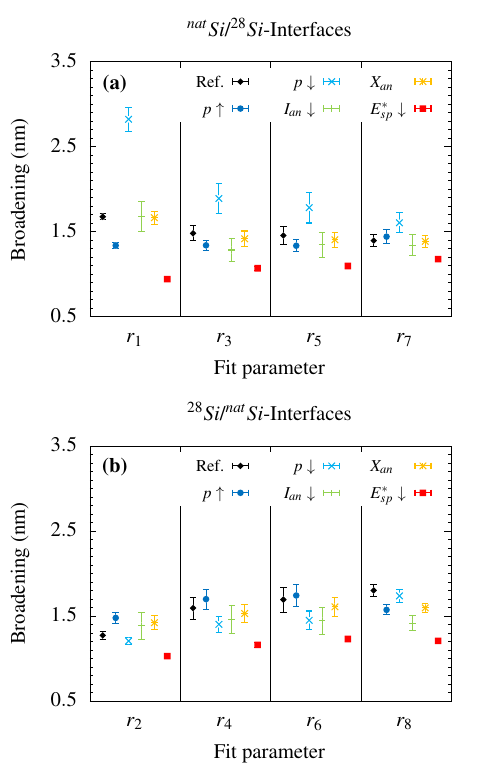} 
    \caption{Graphical presentation of the parameters $r_i$, that describe the broadening of the interfaces, determined by fitting Eq. \ref{Eqn:R2182_Fit-Fkt} to the experimental profiles of the test structure. (a): broadening of the $^{\textit{nat}}$Si/$^{28}$Si interfaces; (b): broadening of the $^{28}$Si/$^{\textit{nat}}$Si interfaces. The error bars indicate the uncertainty of the fit. For the filled squares, the error bars are the same size or smaller than the symbols.}
    \label{Fig:R2182_Broadening}
\end{figure}

The results for the fit parameters $r_i$ that describe the interfacial broadening of sample \#1 are graphically shown in Fig. \ref{Fig:R2182_Broadening}. The broadening of the topmost Si/$^{28}$Si interfaces (Fig. \ref{Fig:R2182_Broadening} (a)) was determined to be slightly smaller than the broadening of the bottom $^{28}$Si/Si interfaces (Fig. \ref{Fig:R2182_Broadening} (b)), i.e. the topmost Si/$^{28}$Si interfaces seem slightly sharper than the bottom $^{28}$Si/Si interfaces. Note that same symbols appear higher on the broadening scale in Fig. \ref{Fig:R2182_Broadening} (a) than in Fig. \ref{Fig:R2182_Broadening} (b). For all measurements performed using a sputtering energy of $\SI{500}{\electronvolt}$, broadening of deeper Si/$^{28}$Si interfaces was determined to be constant within the uncertainty of the fit. However, the uppermost Si/$^{28}$Si interface ($r_{1}$) appears more broadened (e.g. crosses in Fig. \ref{Fig:R2182_Broadening}), due to transient changes in the sputtering and ionization yield, which occur before the sputtering equilibrium has been established. In the case of a lowered flood gas pressure ($p \downarrow$) even deeper interfaces are influenced by changing ionization yields, resulting in a strong broadening (note the decreasing trend of the crosses in Fig. \ref{Fig:R2182_Broadening} (a). For the measurement performed using a sputtering energy of $\SI{250}{\electronvolt}$ ($E_{sp}^{\ast} \downarrow$), the broadening increases slightly with depth (note the increasing trend of the filled squares in Fig. \ref{Fig:R2182_Broadening} (a) and (b)). In this case, the sputtering equilibrium is reached before the topmost Si layer has been completely sputtered. 

The broadening of the $^{28}$Si/Si interfaces (Fig. \ref{Fig:R2182_Broadening} (b)) seems to be slightly increased with depth, using a sputtering energy of $\SI{500}{\electronvolt}$ as well as $\SI{250}{\electronvolt}$. However, a sputtering energy of $\SI{250}{\electronvolt}$ generally leads to the smallest broadening, both for the Si/$^{28}$Si interfaces and for the $^{28}$Si/Si interfaces (note that filled squares are the lowest in Fig. \ref{Fig:R2182_Broadening} (a) and (b)). The error bars indicate that reduced instrumental broadening correlates with higher fitting accuracy. 

A summary of the average changes in the interfacial broadening is given in Tab. \ref{Tab:R2182_SputterRateRatios_Broadening}. Reducing the partial pressure of the flood gas ($p \downarrow$) led to a significantly increased broadening of the Si/$^{28}$Si interfaces (\SI{33}{\%}), resulting in an average increase of the broadening parameter $r$ by about \SI{12}{\%}. Increasing the partial pressure ($p \uparrow$) or using cluster ions for analysis ($X_{an}$) had no significant impact, while reducing the primary ion current ($I_{an} \downarrow$) led to a slightly less broadened profile. The greatest impact has the sputtering energy: while sputtering with $\SI{250}{\electronvolt}$ ($E_{sp}^{\ast} \downarrow$), every interface appears significantly steeper than under reference conditions, resulting in an average reduction of the broadening parameter $r$ by about \SI{27}{\%}. This improvement in broadening coupled with the high fitting accuracy suggests that the profile $E_{sp}^{\ast} \downarrow$ is closest to the true profile of the test structure.

\begin{figure}
    \includegraphics{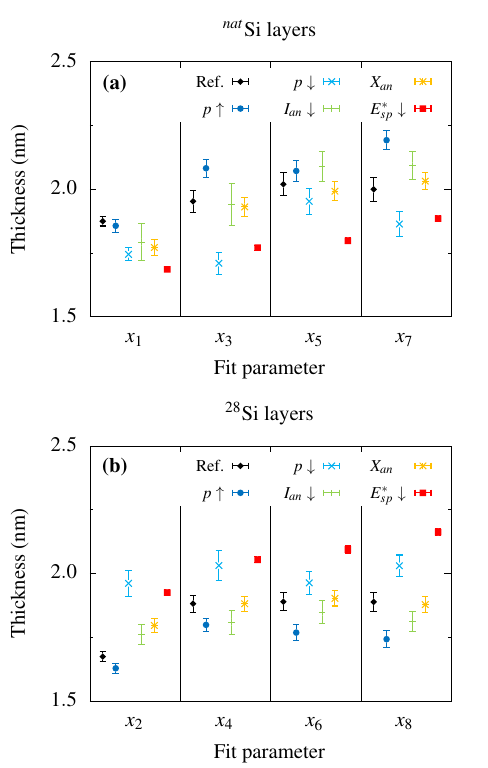} 
    \caption{Graphical presentation of the parameters $x_i$, that describe the thickness of the $^{\textit{nat}}$Si ($x_i$ with $i=1,3,5,7$) and the $^{28}$Si ($x_i$ with $i=2,4,6,8$) layers, determined by fitting Eq. \ref{Eqn:R2182_Fit-Fkt} to the experimental profiles of the test structure. The error bars indicate the uncertainty of the fit. For the filled squares, the error bars are the same size or smaller than the symbols.}
    \label{Fig:R2182_Thickness}
\end{figure}

The fit parameters $x_i$ that describe the layer thickness of sample \#1 are graphically shown in Fig. \ref{Fig:R2182_Thickness}. In the case the primary ion current is reduced ($I_{an} \downarrow$), the primary ion species is changed ($X_{an}$) and the gas pressure of the flood gas is increased ($p \uparrow$), the $^{\textit{nat}}$Si layers were determined to be slightly thicker than the $^{28}$Si layers (the respective symbols appear higher on the thickness scale in Fig. \ref{Fig:R2182_Thickness} (a) than in (b)). In the case the flood gas pressure is reduced ($p \downarrow$) and the sputtering energy is reduced ($E^{\ast}$ $\downarrow$), the $^{\textit{nat}}$Si layers were determined to be slightly smaller than the $^{28}$Si layers (the respective symbols appear lower on the thickness scale in Fig. \ref{Fig:R2182_Thickness} (a) than in (b)). Error bars indicate that the uncertainty in the determination of the layer thickness decreases with profile broadening. Thus, the profile $E_{sp}^{\ast} \downarrow$ revealed the most reliable layer thicknesses that deviate by less than $\SI{0.5}{\nano\metre}$ from the nominal thickness of $\SI{2}{\nano\metre}$. This confirms that the nominal thicknesses were repeatedly achieved during MBE.

\subsubsection{Best sputter and analysis conditions}
The profile $E_{sp}^{\ast} \downarrow$ (Fig. \ref{Fig:R2182_SputterEnergy} (b)) turned out to be closest to the true profile of sample \#1. Topography analysis by means of AFM revealed a root-mean-square roughness of \SI{0.4}{\nano\metre} on the initial surface and \SI{0.5}{\nano\metre} in the ToF-SIMS crater. The slight tendency towards roughening, may explain the slight increase in broadening with depth that was observed in the filled squares in Fig. \ref{Fig:R2182_Broadening} (a) and (b).

\subsection{Depth profiling of a quantum well heterostructure}
\begin{figure}
    \includegraphics{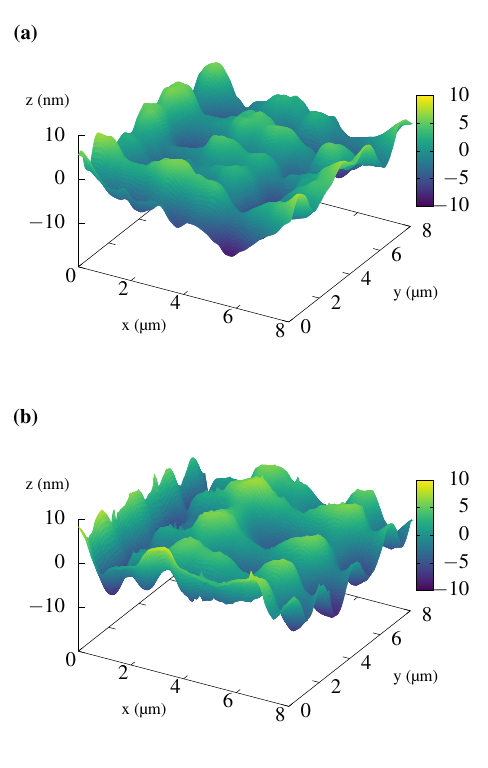} 
    \caption{Topography of a $\SI{8}{\micro\metre} \times \SI{8}{\micro\metre}$ area on the initial surface (a) and in the crater resulting from the ToF-SIMS analysis (b) of sample \#2 analyzed by means of atomic force microscopy. Surface corrugation resulting from strain relaxation via dislocations manifests in regular waves in two perpendicular crystal directions.}
    \label{Fig:R2159_AFM}
\end{figure}

In contrast to the SIMS analysis of homostructures such as the isotopically modulated Si sample \#1, the analysis of semiconductor heterostructures can be affected by the topography of the sample. This, in particular, becomes relevant for semiconductor layer structures with lattice parameters that differ from layer to layer. This lattice mismatch causes internal strain during epitaxial growth that can relax by the formation of dislocations. These misfit dislocations result in corrugated surfaces with a roughness of several nanometers \cite{Kasper2004}.
Prior to SIMS analysis, the topography of sample \#2 was investigated by means of AFM. Fig. \ref{Fig:R2159_AFM} (a) shows the topography of a$\SI{8}{\micro\metre} \times \SI{8}{\micro\metre}$ reference area on the initial surface. The regularly wavy structure of the surface indicates surface corrugation. The root-mean-square roughness was determined to be \SI{3.2}{\nano\metre}. Therefore, ultimate lowered sputter energy on cost of measurement time are not suitable in this case because the surface roughness limits the depth resolution. Taking the results from the parameter variation study on sample \#1 into account, the instrumental setup used for reference purposes was chosen for the depth profile analysis of sample \#2. These settings provide a sufficient depth resolution in view of the surface roughness and, in case SIMS analyses are performed commercially, the reduced measurement time due to the higher sputter energy limits the costs for the analysis.

\begin{figure}
    \includegraphics{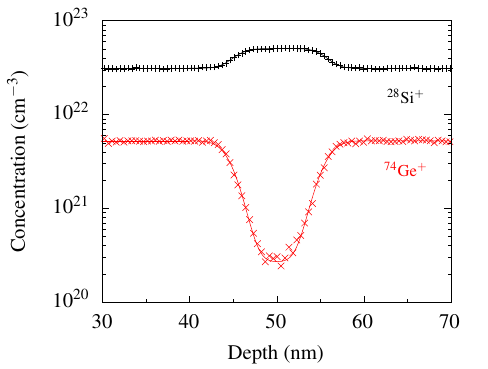} 
    \caption{SIMS concentration-depth profiles of $^{28}$Si and $^{74}$Ge in the quantum well region of the Si$_{0.7}$Ge$_{0.3}$/$^{28}$Si/Si$_{0.7}$Ge$_{0.3}$ heterostructure. Measurement conditions: O$_2^{+}$, \SI{500}{\electronvolt}, \SI{40}{\nano\ampere}, $\SI{300}{\micro\metre} \times \SI{300}{\micro\metre}$ sputtering; Bi$_1^{+}$, \SI{15}{\kilo\electronvolt}, \SI{0.25}{\pico\ampere}, $\SI{75}{\micro\metre} \times \SI{75}{\micro\metre}$ analysis; O$_2$, \SI{2E-6}{\milli\bar} flooding. Plus-signs mark Si data and crosses mark Ge data. Every 10th data point is shown for clarity. The solid lines represent best fits based on Eq. \ref{Eqn:R2159_Fit-Fkt} taking sputter broadening effects into account.}
    \label{Fig:R2159_Profiles}
\end{figure}

Fig. \ref{Fig:R2159_Profiles} shows the measured concentration-depth profiles of $^{28}$Si and $^{74}$Ge in the quantum well region. In general, the structure is well resolved. The thickness of the quantum well was determined to be \SI{10.5\pm0.2}{\nano\metre} which corresponds to STEM data for this sample \cite{Klos2024}. In the absence of a suitable standard sample, a relative sensitivity factor (RSF) for Ge in Si could not be determined. Assuming constant sputter and ionization yields through the whole structure, the $^{74}$Ge concentration was determined to be \SI{2.7\pm0.1E+20}{\centi\metre^{-3}} in the $^{28}$Si quantum well (fit parameter $C_2$). Taking the isotopic abundance of Ge into account, the residual Ge concentration within the $^{28}$Si layer was calculated to \SI{7.4E+20}{\centi\metre^{-3}} (\SI{1.5}{at.\%}). The residual concentration of $^{29}$Si in the MBE source material is $41$\,ppm (\SI{0.0041}{at.\%}) according to the supplier which is confirmed by means of APT analysis of the quantum well structure that yields a $^{29}$Si concentration of \SI{50\pm20}{ppm} (\SI{0.005\pm0.002}{at.\%}) in the $^{28}$Si quantum well \cite{Klos2024}. The APT measurement yields a Ge concentration of \SI{0.3}{at.\%} in the quantum well \cite{Klos2024}. Obviously, the Ge concentration in the quantum well is overestimated by means of ToF-SIMS. This might be caused by the matrix effect, i.e. the intensities of ($^{74}$Ge) secondary ions sputtered from Si$_{0.7}$Ge$_{0.3}$ and from residual $^{74}$Ge in $^{28}$Si are different.

The fit parameters $r_i$, that describe the interfacial broadening, determined by fitting Eq. \ref{Eqn:R2159_Fit-Fkt} to the $^{28}$Si data and the Ge data respectively, are summarized in Tab. \ref{Tab:R2159_Broadening}.  
\begin{table}
    \centering
    \caption{Fit parameters $r_i$, that describe the interfacial broadening at the top interface ($i=1$) and at the bottom interface ($i=2$) of the quantum well heterostructure, determined by fitting Eq. \ref{Eqn:R2159_Fit-Fkt} to the $^{28}$Si data and the $^{74}$Ge data respectively.}
    \begin{ruledtabular}
    \begin{tabular}{lcc}
    \multicolumn{1}{l}{} & \multicolumn{1}{c}{$r_{1}$ ($\si{\nano\metre}$)} & \multicolumn{1}{c}{$r_{2}$ ($\si{\nano\metre}$)} \\
    \hline
    $^{28}$Si & $1.9 \pm 0.1$ & $1.8 \pm 0.1$  \\
    $^{74}$Ge & $2.1 \pm 0.1$ & $2.2 \pm 0.1$  \\
    \end{tabular}
    \end{ruledtabular}
    \label{Tab:R2159_Broadening}
\end{table}
The broadening at the top interface (Si$_{0.7}$Ge$_{0.3}$/$^{28}$Si) equals the broadening at the bottom interface ($^{28}$Si/Si$_{0.7}$Ge$_{0.3}$) within the uncertainty of the fit. However, the Ge profile leads to a tendentious higher broadening than the $^{28}$Si profile. This likely indicates a surface that is preferred terminated with Ge during growth \, \cite{Kube2010}.

To ensure that sputtering has not induced additional roughness, the topography analysis by means of AFM was repeated in the crater. As shown in Fig. \ref{Fig:R2159_AFM} (b) surface corrugation can be observed manifesting in regular waves on a \si{\micro\metre} scale in two perpendicular crystal directions. The root-mean-square roughness was determined to be \SI{3.2}{\nano\metre}. This is the same value as determined for the reference position on the initial surface, i.e. sputtering has not changed the surface roughness.

\section{Conclusion}
ToF-SIMS has proven to be a suitable characterization method to determine isotopic concentration-depth profiles of semiconductor heterostructures. It has been shown that instruments used for routine analysis can also achieve a depth resolution below \SI{1}{\nano\metre}, as long as the topography of the analyzed samples does not limit the depth resolution itself. When analyzing a test homostructure consisting of ultra-thin $^{\textit{nat}}$Si/$^{28}$Si bilayers of only \SI{2}{\nano\metre} in thickness, the instrumental setting was optimized in favor of depth resolution, whereby, as expected, sputtering energy turned out to be one key parameter. It has been demonstrated that the consideration of the sputter rate ratio, i.e. the amount of material sputtered by both, the sputter beam and the analysis beam, plays a crucial role. In comparison to an instrumental setting used for routine analysis, the profile broadening was reduced by \SI{27}{\%} at the expense of measurement time, which was extended by the factor $14$ in this study. Based on the results of this study, it is conceivable that a further reduction in sputtering energy could lead to a further reduction in instrumental broadening. However, $\SI{250}{\electronvolt}$ ($E_{sp}^{\ast} \downarrow$) was the lowest possible sputtering energy that could be achieved with the device used, limited by the electronics that control the high voltage of the sputtering ion beam. Furthermore, it was shown that a SiGe/Si/SiGe heterostructure with a surface corrugation of several \si{\nano\metre} caused by the lattice mismatch between SiGe and Si can be successfully resolved. The thickness of the quantum well layer of this structure was determined to be \SI{10.5\pm0.2}{\nano\metre} in  correspondence to STEM data.

\begin{acknowledgments}
This work was funded by the German Research Foundation (Deutsche Forschungsgemeinschaft, DFG) within the projects 421769186 (SCHR 1404/5-1), 289786932 (SCHR 1404/3-2) and 289786932 (BO 3140/4-2). The research is also part of the Munich Quantum Valley, which is supported by the Bavarian state government with funds from the Hightech Agenda Bavaria.

Thanks to Guido Winkler for measurement support and helpful discussions.
\end{acknowledgments}

\section*{Data Availability}
The data that support the findings of this study are available from the corresponding author upon reasonable request.

\section*{Author Declarations}
The authors have no conflicts to disclose.

\bibliography{references}

\end{document}